\documentclass[aps,prl,reprint,groupedaddress,twocolumn,floatfix]{revtex4}
\usepackage{graphicx}   
\usepackage{amsmath,amssymb,wasysym}
\usepackage[amsmath,thmmarks,framed]{ntheorem}
\usepackage{color}
\usepackage{hyperref}

\begin{document}

\newcommand{\idunno}[0]{\fcolorbox{red}{blue}{\color{white} 666}}
\newcommand{\lolwut}[1]{\fcolorbox{red}{blue}{\color{white} #1}}
\newcommand{\cesium}[0]{$^{133}$Cs}
\newcommand{\caesium}[0]{Cs}
\newcommand{\uW}[0]{\,$\mu$W}
\newcommand{\uS}[0]{\,$\mu$s}
\newcommand{\fthz}[0]{\,fT/$\sqrt{\text{Hz}}$}
\newcommand{\pthz}[0]{\,pT/$\sqrt{\text{Hz}}$}
\newcommand{\brf}[0]{$\vec{B}_{\text{rf}}$}
\newcommand{\brfm}[0]{\vec{B}_{\text{rf}}}
\newcommand{\bmain}[0]{$\vec{B}_0$}
\newcommand{\temp}[1]{$#1^{\circ{}}\!\text{C}$}
\newcommand{\F}[1]{$F=#1$}
\newcommand{\FF}[2]{$F=#1 \rightarrow F'=#2$}

\newcommand{\xz}[1]{\fcolorbox{red}{blue}{\color{white}#1}}
\newcommand{\xzcite}[0]{\xz{?}}
\newcommand{\lolrly}[0]{\xz{RLY?}}
\newcommand{\axis}[1]{\textsf{#1}}
\newcommand{\parVec}[2]{#1\,$\parallel$\,#2}
\newcommand{\perpVec}[2]{#1\,$\perp$\,#2}
\newcommand{\kprobe}[0]{$\vec{k}_{pr}$}
\newcommand{\kpump}[0]{$\vec{k}_{pmp}$}
\newcommand{\kls}[0]{$\vec{k}_{LS}$}
\newcommand{\hlc}[2][yellow]{ {\sethlcolor{#1} \hl{#2}} }
\newcommand{\myMaxColWidth}{4cm}

\newcommand{\labnoiseSTEP}[0]{$40$\fthz}
\newcommand{\labnoiseSTEPLS}[0]{$55$\fthz}
\newcommand{\labnoiseLSD}[0]{$40$\fthz}
\newcommand{\shotnoise}[0]{$1.7$\fthz}
\newcommand{\SNLKitching}[0]{$2$\pthz}
\newcommand{\pumpnoise}[0]{$3.3$\fthz}
\newcommand{\shotnoiseSync}[0]{$1.9$\fthz}
\newcommand{\lsmodfreq}[0]{$0.8$\,Hz}
\newcommand{\fgmodfreq}[0]{$0.5$\,Hz}
\newcommand{\samplelength}[0]{$40$\,s}
\newcommand{\samplingrate}[0]{\lolwut{$200$\,sps}}
\newcommand{\interferencePeriod}[0]{$60$\,GHz}
\newcommand{\lspower}[0]{$5.5$\uW{}}

\newcommand{\resonatorLength}[0]{$1.7$\,mm}
\title{Light shift averaging in paraffin-coated alkali vapor cells}

\author{Elena Zhivun}
\affiliation{Department of Physics, University of California, Berkeley, California 94720-7300, USA}
\author{Arne Wickenbrock}
\affiliation{Johannes Gutenberg-University Mainz, 55128 Mainz, Germany}
\email[]{wickenbr@uni-mainz.de}
\author{Julia Sudyka}
\affiliation{Marian Smoluchowski Institute of Physics, \L ojasiewicza 11, 30-348 Krak\'ow, Poland} 
\author{Brian Patton\footnote{Present address: AOSense, Inc., Sunnyvale, CA 94085,
USA.}}
\affiliation{Department of Physics, University of California, Berkeley, CA 94720-7300 and\\
Physik-Department, Technische Universit\"{a}t M\"{u}nchen, 85748 Garching, Germany}
\author{Szymon Pustelny}
\affiliation{Marian Smoluchowski Institute of Physics, \L ojasiewicza 11, 30-348 Krak\'ow, Poland} 
\author{Dmitry Budker}
\affiliation{Helmholtz Institute, Johannes Gutenberg-University, 55099 Mainz, Germany\\ Department of Physics, University of California, Berkeley, CA 94720-7300 and \\ Nuclear Science Division, Lawrence Berkeley National Laboratory, Berkeley, CA 94720}
\date{\today}

\begin{abstract}
Light shifts are an important source of noise and systematics in optically pumped magnetometers.
We demonstrate that the long spin coherence time in paraffin-coated cells leads to spatial averaging of the light shifts over the entire cell volume. This renders the averaged light shift independent, under certain approximations, of the light-intensity distribution within the sensor cell. 
These results and the underlying mechanism can be extended to other spatially varying phenomena in anti-relaxation-coated cells with long coherence times.
\end{abstract}

\pacs{}

\maketitle
%


\section{Introduction}\label{introduction}
Light shifts in alkali atoms have been researched since 1960s \cite{ArditiLightShifts, HapperLightShifts}, however the effect of laser-induced vector light shifts (VLS) in coated cells has been little explored in the literature.
The works considering light-shift effects primarily focus either on transitions relevant for atomic clocks \cite{ArditiHFS, LightShiftCPT, LightShiftCPT2, BrilletMetrologia, OshimaCsBeam, BudkerMicrowave}, magnetometers using buffer-gas cells \cite{WalkerACStarkInSERF, Podvyaznyi, CamparoInhomogeneousLS, SkallaLightShift}, or light shifts induced by broad-spectrum alkali metal lamps \cite{ArditiLightShifts, HapperLightShifts, BulosRealTransitions, CohenTannoudjiZeemanLS}.

In this paper, we discuss the role of VLS in an optical magnetometer exploiting a paraffin-coated alkali-metal vapor cell \cite{Higbie2006, Acosta2006, Patton2012, Bison2006, Bison2015, Seltzer2007, Lucivero2014, Szymek1}.
Unlike uncoated cells, paraffin-coated cells enable the atoms to undergo a number of wall collisions (up to $10^6$~\cite{supercell}) without depolarization.
In this way, thermal atoms sample the entire cell volume during their spin relaxation time and hence become sensitive to average, rather than local magnetic field \cite{SpatialAveraging2006}. 
In this work we demonstrate that the same reasoning can be applied to the VLS in paraffin-coated cells, making it, under realistic assumptions, a function of the total light power averaged over the cell volume, rather than the local intensity.

The experiment consists of a series of VLS measurements as a function of beam diameter and optical frequency in a synchronously pumped (Bell-Bloom \cite{BellBloom, DimaOpticalMagnetometry}) orientation-based \caesium{} magnetometer. The magnetometer utilizes the resonant response of the atoms that occurs when the atoms are exposed to light modulated at a frequency that matches their Larmor-precession frequency $\omega_L$. The magnetic field can then be extracted by measuring the center frequency of the Lorentzian shaped magnetic resonance (MR). We investigate the dependence of the light shift on the beam diameter for a fixed optical power and show that with negligible optical pumping the shift is independent of the size of the laser beam (apart from small systematic contributions).
\section{Model}\label{theory}
To gain understanding of the processes involved in the light shift averaging, we consider a concentric cylindrical cell and a laser beam with radii $R$ and $r$, respectively.
The optical frequency of the light-shift laser is far-detuned from the relevant atomic transitions compared to the Doppler width, and the beam-intensity profile is assumed flat (top hat). In the presence of a magnetic field the atomic magnetic dipoles precess with $\omega_L$. Each individual atom interacting with the light-shift beam acquires a phase advance or retardation $\phi$ in its Larmor precession proportional to the change in the Larmor frequency due to the vector light shift in the beam $\delta \nu$ and the time $t$ spent in it during the spin coherence time $T_2$. The change in the Larmor frequency by a circularly polarized laser beam in the proximity of the D$_2$ transition is given by:
\begin{equation}
\delta \nu \approx \beta \frac{I}{\Delta_{LS}},
\label{equation1}
\end{equation}
where $\Delta_{LS}$ is the frequency detuning of the LS beam with respect to the center of the Voigt profile of the \FF{4}{3,4,5}  transitions, $I$ the intensity of the LS beam, and $\beta$ is a proportionality constant that can be calculated with second-order perturbation theory taking all the relevant transitions and magnetic sub-levels into account (see, for example, \cite{Rauschenbeutel}).

The scaling of the MR center frequency shift can be understood as follows:
assuming that the probability to find an atom in any part of the cell volume is the same, $t$ is proportional to $r^2/R^2$, while the beam intensity $I$ scales as $1/r^2$. 
Since the average phase change acquired by an atom is $\phi \propto I \times t$, it does not depend on $r$.  This average phase increment during $T_2$ can be translated into a average frequency shift of the MR $\delta\nu_{LS}$ by dividing it by the coherence time $T_2$ of the Larmor precession. Since $\phi$ is independent of the beam area, $\delta\nu_{LS}$ stays constant for a given power.

To estimate the LS beam contributions to the width of the MR we have to consider the different stochastic processes involved and their probability density functions. How often an atom interacts with the LS beam follows a Poissonian distribution with mean $n \propto r/R$ and how much phase advance or retardation the atom picks up should be normally distributed around $\phi=n\,\phi_1$ with variance $n\,\delta\phi_1^2$ (where $\phi_1$ and $\delta\phi_1^2$ are the mean and variance per interaction, respectively). For $\sqrt{n}\,\delta\phi_1< n\,\left|\phi_1\right|\ll1$ it can be shown (see, Ref.  \cite{DimaProblems}) that this kind of Larmor precession phase diffusion leads to the previously discussed shift but also to a broadening of the MR. 
The width increase is due to the uncertainty in the number of interactions per atom as well as the variance $\delta\phi_1^2$ of the LS effect for a single interaction. 
If the width of the MR stems from a combination of different phase-diffusion processes the contributions add in quadrature.
\section{Experimental setup}\label{setup}
\begin{figure}[bth]
  \centering
  \includegraphics[width=\columnwidth]{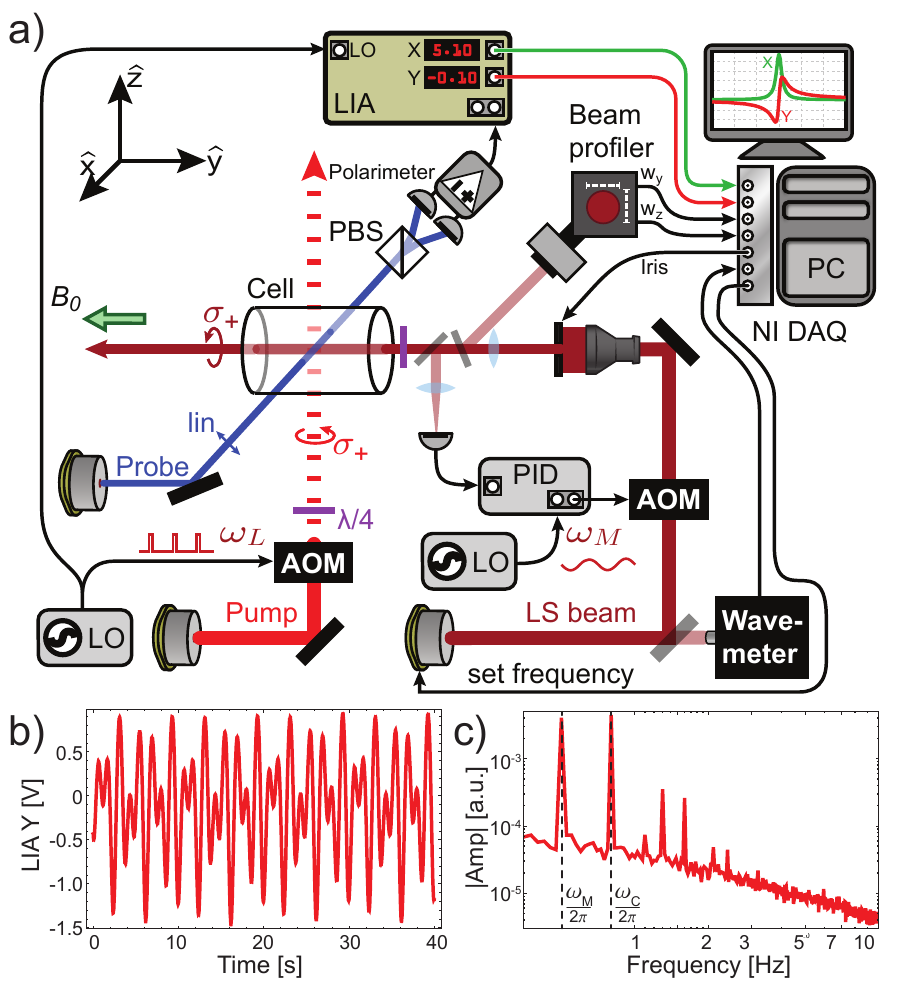}
  \caption{(Color online) a) The experimental setup. The amplitude-modulated,
circularly polarized pump beam propagates along $\hat{z}$ (orthogonally to \parVec{\bmain{}}{\^{y}}).
A local oscillator (LO) pulses the pump intensity via an acousto-optical modulator (AOM) and serves as a reference for a lock-in amplifier (LIA), whose analog output is recorded with a data acquisition card (DAQ) and stored on a computer (PC).
After transmitting through the cell, the linearly polarized probe beam is analyzed with a balanced polarimeter, consisting of a polarizing beam splitter (PBS) and two photodiodes.
The circularly polarized light-shift beam (LS beam) propagates along \bmain{}.
Its diameter is varied with a computer controlled, motorized iris while its time-averaged power is kept constant with an AOM in a feedback loop.
An image of the iris is formed inside the cell and on the beam profiler using a lens system.
The optical frequency of the LS beam is measured with a wavemeter and controlled by the PC.
For noise reduction, we perform synchronous detection of the VLS signal while harmonically modulating the LS beam power at $\omega_M$.
b) shows the recorded time series for the LIA Y output for a single light shift measurement with the simultaneous modulation of the LO frequency and the LS power. 
c) The FFT of the signal in b) shows the calibration peak at $\omega_C/2\pi$ and the LS amplitude at $\omega_M/2\pi$.  }\label{lsd_setup}
\end{figure}

The foundation of the experimental setup (Fig. \ref{lsd_setup}a) is a synchronously pumped Bell-Bloom type magnetometer with an additional laser beam (LS beam) that induces the light shift. 
The magnetic shielding, the sensor cell, and the leading field \bmain{} source are described in Ref.~\cite{vectormag}.
The paraffin-coated \cesium{} cell has a cylindrical shape and measures 50\,mm in length and in diameter. Its longitudinal spin relaxation time is 0.7\,s.
The leading field \bmain{} has a magnitude of $489$\,nT, ($\omega_L = 2\pi\cdot1710.0$\,Hz for the \caesium{} \F{4} manifold), and is parallel to \^{y}.  
The probe light (852\,nm distributed feedback laser (DFB) from Eagleyard, $9.8$\uW) is linearly polarized ($\vec{E}\| \hat{y}$) and propagates orthogonally to \parVec{\bmain{}}{\^{y}} (wave vector \parVec{\kprobe{}}{\^{x}}).
Its optical frequency is stabilized using a dichroic atomic vapor laser lock (DAVLL, not shown in Fig. \ref{lsd_setup}a)~\cite{DAVLL1, DAVLL2} close to the \F{4} manifold. The detuning and intensity of the probe beam are optimized for the largest signal with minimal power broadening.
The circularly polarized pump light (852\,nm DFB, $16.8$\uW{} time averaged) is injected into the cell orthogonally to \bmain{} (\parVec{\kpump{}}{\^{z}}).
Its power is pulsed by an AOM at the Larmor frequency with a 3.4\% duty cycle.
The beam is routed by a single-mode polarizing fiber (Fibercore HB830Z) with cleanup polarizers at the output.
The pump frequency is locked on resonance with the \FF{3}{4} transition by a DAVLL (not shown in Fig. \ref{lsd_setup}).
This produces atomic polarization (orientation) in the $F=4$ manifold by optically pumping while simultaneously depopulating the \F{3} manifold. The large detuning from the probed \F{4} manifold also minimizes power broadening and light shifts caused by the pump~\cite{Szymek}. 
The LS beam is circularly polarized (895\,nm DFB, $5.5$\uW{}  time averaged) and propagates parallel to \bmain{} so that the vector light shift adds linearly to the Zeeman shift due to \bmain{}.
The beam power is actively stabilized with an AOM in a feedback loop.
To change the size of the beam  and ensure a homogeneous intensity distribution, the initially Gaussian beam is expanded by a telescope to a large diameter and the central part is picked out by a computer controlled iris with variable aperture. The image of the iris is then formed by a lens system inside the cell and onto a beam profiler (Coherent LaserCam RS-170). For each beam size an image is taken and saved for later analysis.
The pick-off for power stabilization and beam profiling is provided by an uncoated wedged beam splitter positioned after the imaging optics outside of the shields just followed by polarizing elements in front of the cell. The optical frequency of the LS beam is actively controlled by a wavemeter ($\mathring{\textrm{A}}$ngstrom/HighFinesse {WS-7}) in a feedback loop.

As the probe light propagates through the polarized atomic vapor, it experiences magneto-optical rotation~\cite{DimaAlignment} which causes a modulation of the probe's polarization direction at the Larmor frequency.
The optical signal is detected with a balanced polarimeter connected to a transimpedance amplifier and demodulated with a lock-in amplifier (LIA) at the modulation frequency.

A signal generator (BNC 645) controls the pump pulse frequency and serves as a local oscillator (LO) for the LIA (SR 830).
The phase of the LIA is adjusted to produce absorptive and dispersive signals in the \textsf{X} and \textsf{Y} LIA channels respectively, when the pump-pulse repetition frequency is scanned across the MR.
The \textsf{X} and \textsf{Y} signals are captured with a data acquisition system (NI DAQ 6353) and analyzed with a computer (PC) that controls the experiment.
Slow drifts in the the MR center frequency and width pose a challenge for our measurement, as they cause all data points to have a slightly different scaling between the measured signal and the actual light shift. 
To mitigate this, we simultaneously harmonically modulate the LS beam power and the pump pulse frequency (in the vicinity of the resonance) at different frequencies. Both signals cause a linear response in the dispersive channel of the lock-in. The latter modulation enables us to continuously monitor the dispersive slope of the MR throughout the experiment since the modulation amplitude is known. We effectively compare the modulation due to the LS beam with a modulation of known amplitude. This way the response of the cell is calibrated for each data point.
The modulation frequencies of the LS beam power ($\omega_M=2\pi\cdot$\lsmodfreq{}) and the pump pulse frequency ($\omega_C=2\pi\cdot$\fgmodfreq{}) are chosen to be smaller than the MR linewidth ($3.2$\,Hz) to avoid low-pass filtering.

For each data point, we record a \samplelength{} sample of \textsf{X} and \textsf{Y} channels of the LIA. An example can be seen in Fig. \ref{lsd_setup}b.
The signals are Fourier transformed (Fig. \ref{lsd_setup}c) and processed to extract the MR linewidth, amplitude and RMS Larmor frequency deviation caused by the LS beam.
Each sample has an integer number of periods of both LS and LO modulations to minimize spectral leakage.
\section{Results}\label{results}
\begin{figure}[bth]
  \centering
  \includegraphics[width=\columnwidth]{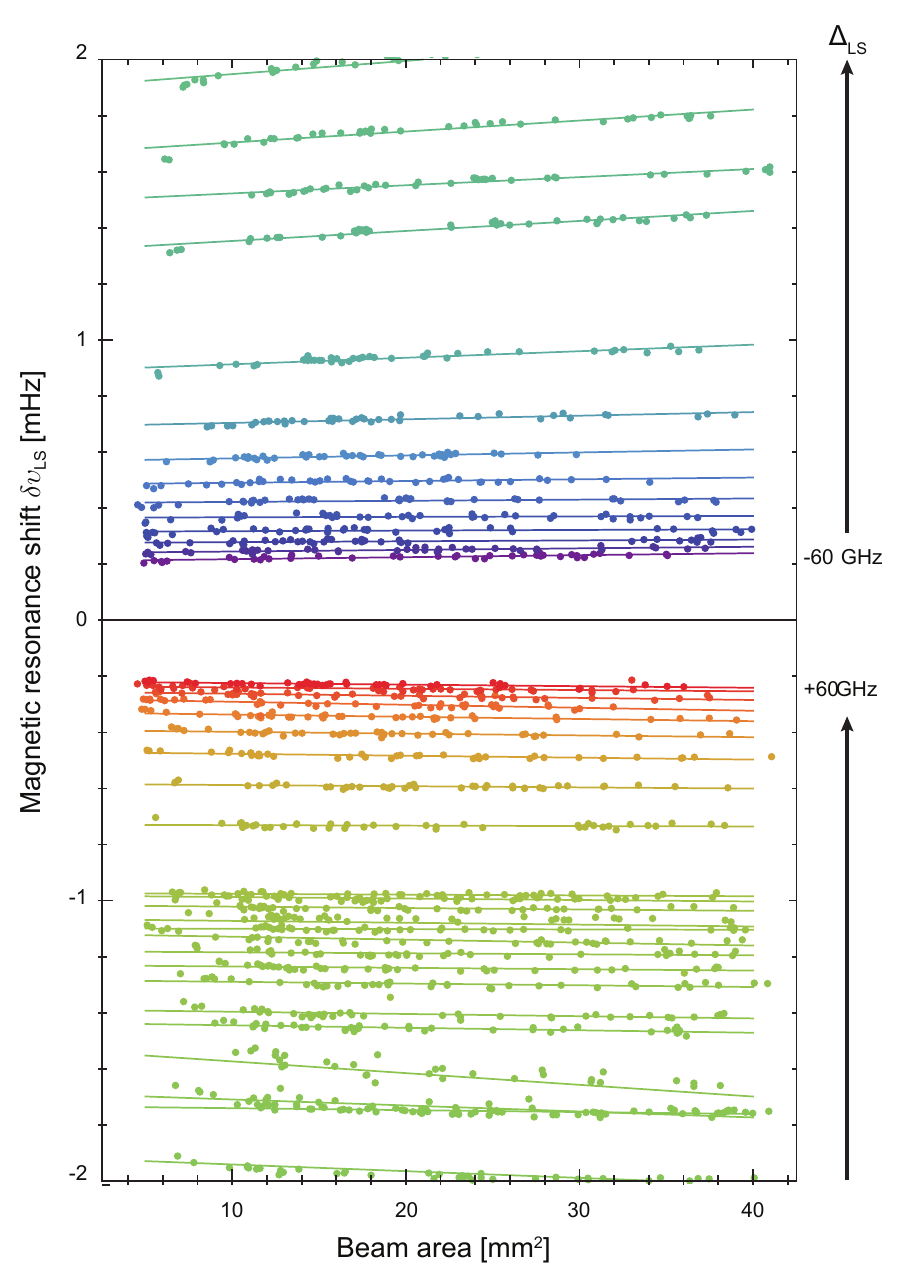}
  \caption{(Color online) Change of the magnetic resonance center frequency as a function of the light-shift beam area for different LS beam detunings and a constant beam power.
  The complete data include detunings from $-60$\,GHz to $+60$\,GHz with respect to the \cesium{} D2 \F{4} transitions.
  While the beam area and therefore the beam intensity is modified by an order of magnitude, the MR center frequency changes are on average 3\% and are of technical origin as explained in the text.
  Different colors represent distinct optical frequencies of the LS beam, the detuning is indicated by the arrows on the right.
  The data points are represented by circles, and the lines are linear fits to the datasets.
  The fit parameter are average light shift and light shift change per unit area change. }\label{mr_shift_vs_beam_area}
\end{figure}

\begin{figure}[bth]
  \centering
  \includegraphics[width=0.95\columnwidth]{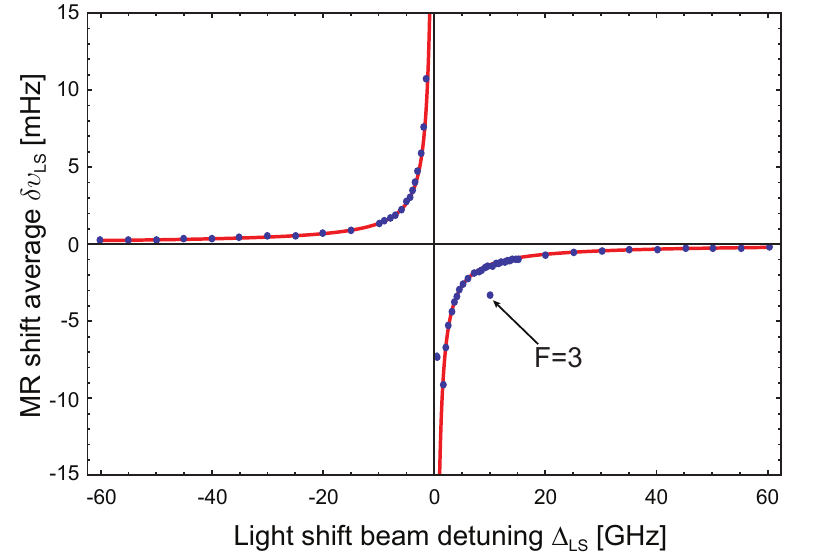}
  \caption{(Color online) Average vector light shift dependence on the optical frequency as derived from the fits to the data displayed in Fig.~\ref{mr_shift_vs_beam_area}.  The error bars are hidden within the points since the average ratio between data value and error is 140. The red curve shows a fit to the data with $\propto 1/{\Delta_{LS}}$.}
\label{mr_shift_vs_f0}
\end{figure}

\begin{figure}[bth]
  \centering
  \includegraphics[width=0.95\columnwidth]{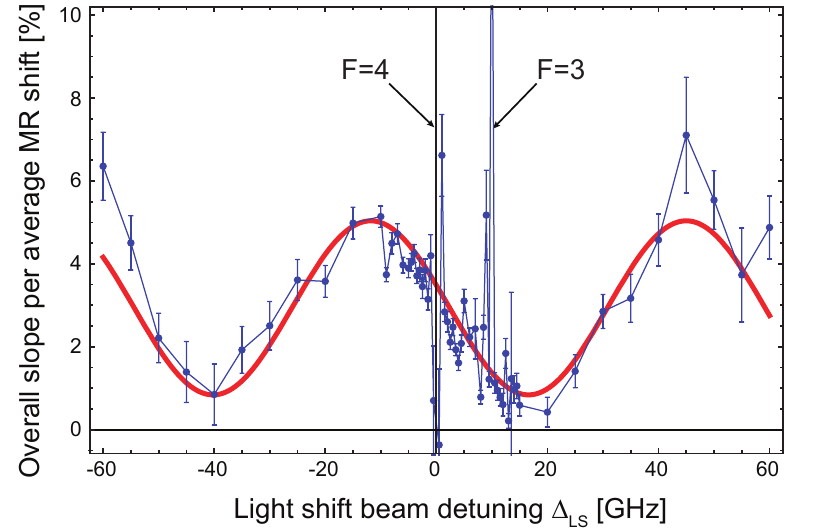}
  \caption{(Color online) Maximum light shift change with beam area from Fig. \ref{mr_shift_vs_beam_area}  divided by the average light-shift as a function of the optical light shift frequency.
  The red line is a sinusoidal fit revealing an etalon effect potentially in the cell wall. }\label{mr_shift_error_vs_f0}
\end{figure}

To see whether the VLS is averaged by the atoms in a paraffin-coated cell, we measure the dependence of the MR center shift as a function of the LS beam area for a given total power (\lspower{}).
The average LS intensity is adjusted to keep the total power constant as the beam diameter changes.
The procedure is repeated for different  optical frequencies of the LS beam, and each data set is fitted with a linear function.
The fit results for different LS beam optical frequencies are presented in Fig. \ref{mr_shift_vs_beam_area}.

The average light shift was calculated as an average Larmor frequency change over the data points with different diameters and a given optical frequency.
As shown in Eq. \ref{equation1} the average light shift scales as $1/\Delta_{
LS}$ for large frequency detunings. This dependence is clearly visible in Fig. \ref{mr_shift_vs_f0}.

The outlier point appearing at $+9$\,GHz in Fig. \ref{mr_shift_vs_f0} and \ref{mr_shift_error_vs_f0} corresponds to the \FF{3}{2,3,4} transitions.
This deviation from the theoretical scaling of the VLS and the slope is a result of optical pumping by the LS beam. At this frequency the LS beam acts as an additional pump producing orientation in the \F{4} manifold with effects on the MR width and amplitude. The absorption in the vapor also causes an effective reduction of the LS beam intensity. 
However, since it is an on-resonant effect a detailed analysis is beyond the scope of this paper.
Throughout the experiment the maximum MR center frequency shift was below $100$\,mHz ($30$\,pT), which is  well within the MR resonance width. 
Signal-to-noise ratio for each data point in Fig. \ref{mr_shift_vs_f0} was on average $140$ and for large detunings the measurement error was below $1$\,fT.

The dependence of the VLS on the beam diameter for each optical frequency is presented in Fig. \ref{mr_shift_error_vs_f0}.
The plot is the result of normalizing the light shift change per area change (Hz/mm$^2$) by the average light shift at each optical frequency, i.e. the ratio of the slope in Fig. \ref{mr_shift_vs_beam_area} and the average light shift scaled by the maximum area change.
A change of the beam area (and therefore of the intensity) by a factor of eight results in an average light shift change on the order of 3\%. A likely explanation of this small variation is a systematic effect related to the power stabilization of the LS beam. There seems to be a small diameter dependent difference in power between the cell and the stabilization photodiode. This can have multiple reasons, e.g., clipping of the beam or an angle dependent sensitivity of the photodiode.

In addition to the average change, a sinusoidal variation with the optical frequency detuning is visible.
This oscillation pattern appears to be the result of an etaloning effect in the vapor-cell windows. The modulation period of \interferencePeriod{} corresponds to a glass-resonator length of \resonatorLength{}, which is consistent with the thickness of the cell windows. 
To further verify this, we measured the LS beam power transmission as a function of beam diameter for different LS beam frequencies, which revealed the same pattern.

In a separate set of similar experiments (not discussed here) we tried to observe the LS beam induced broadening of the MR width. Even with much longer averaging no broadening was detected. This is not surprising however given the values of $\delta\nu_{LS}$ are much smaller than the MR width. We expect the additional contribution due to $\delta\nu_{LS}$ to be on the order of $\delta\nu_{LS}/\sqrt{n}$, which should be added to the MR width in quadrature (Sec. II). The resulting maximum increase in MR width is less than 10\,ppm, which was experimentally inaccessible.

\section{Conclusion}\label{conclusion}
In conclusion, we investigated how the vector light shift exerted by a circularly polarized laser beam on atoms in a paraffin-coated cell depends on the beam area.
Theoretical estimation suggests that for a given beam power the overall light shift should be independent of the beam area, as long as the thermal atoms adequately sample the entire cell volume during the spin relaxation time.
We experimentally verified that the vector light shift in a coated cell depends nearly exclusively on the total beam power and not on the beam area.
With a factor of eight change in the beam area, the light shift changes by less than 3\%.
The residual dependence on the area can be explained by frequency and diameter dependent transmission of the LS beam in the optical elements between the atoms and the intensity-stabilization photodiode.
The magnetic resonance broadening due to the variance in the number of passes through the light-shift beam or due to the variance in the time spent in the beam per pass was below the experimental sensitivity.
These results are important for modern magnetic sensors that make use of auxiliary fictitious fields~\cite{vectormag} and can be extended to other spatially averaged quantities in cells with long coherence times.

\begin{acknowledgments}
We would like to thank Larry~R.~Hunter for  helpful input and comments. This research was supported in part by the National Science Foundation under award CHE-1308381 and by the NGA NURI program. JS and SP would like to acknowledge support from  Marie Curie programme, FP7 ``Coherent optics sensors for medical applications-COSMA'' (PIRSES-GA-2012-295264), and the National Centre for Research and Development within the Leader program.
\end{acknowledgments}.
\bibliography{literature}

\newcommand{\noopsort}[1]{} \newcommand{\printfirst}[2]{#1}
  \newcommand{\singleletter}[1]{#1} \newcommand{\switchargs}[2]{#2#1}
\begin{thebibliography}{33}
\expandafter\ifx\csname natexlab\endcsname\relax\def\natexlab#1{#1}\fi
\expandafter\ifx\csname bibnamefont\endcsname\relax
  \def\bibnamefont#1{#1}\fi
\expandafter\ifx\csname bibfnamefont\endcsname\relax
  \def\bibfnamefont#1{#1}\fi
\expandafter\ifx\csname citenamefont\endcsname\relax
  \def\citenamefont#1{#1}\fi
\expandafter\ifx\csname url\endcsname\relax
  \def\url#1{\texttt{#1}}\fi
\expandafter\ifx\csname urlprefix\endcsname\relax\def\urlprefix{URL }\fi
\providecommand{\bibinfo}[2]{#2}
\providecommand{\eprint}[2][]{\url{#2}}

\bibitem[{\citenamefont{Arditi and Carver}(1961)}]{ArditiLightShifts}
\bibinfo{author}{\bibfnamefont{M.}~\bibnamefont{Arditi}} \bibnamefont{and}
  \bibinfo{author}{\bibfnamefont{T.~R.} \bibnamefont{Carver}},
  \bibinfo{journal}{Phys. Rev.} \textbf{\bibinfo{volume}{124}},
  \bibinfo{pages}{800} (\bibinfo{year}{1961}),
  \urlprefix\url{http://link.aps.org/doi/10.1103/PhysRev.124.800}.

\bibitem[{\citenamefont{Mathur et~al.}(1968)\citenamefont{Mathur, Tang, and
  Happer}}]{HapperLightShifts}
\bibinfo{author}{\bibfnamefont{B.~S.} \bibnamefont{Mathur}},
  \bibinfo{author}{\bibfnamefont{H.}~\bibnamefont{Tang}}, \bibnamefont{and}
  \bibinfo{author}{\bibfnamefont{W.}~\bibnamefont{Happer}},
  \bibinfo{journal}{Phys. Rev.} \textbf{\bibinfo{volume}{171}},
  \bibinfo{pages}{11} (\bibinfo{year}{1968}),
  \urlprefix\url{http://link.aps.org/doi/10.1103/PhysRev.171.11}.

\bibitem[{\citenamefont{Arditi and Picque}(1975)}]{ArditiHFS}
\bibinfo{author}{\bibfnamefont{M.}~\bibnamefont{Arditi}} \bibnamefont{and}
  \bibinfo{author}{\bibfnamefont{J.~L.} \bibnamefont{Picque}},
  \bibinfo{journal}{Journal of Physics B: Atomic and Molecular Physics}
  \textbf{\bibinfo{volume}{8}}, \bibinfo{pages}{L331} (\bibinfo{year}{1975}),
  \urlprefix\url{http://stacks.iop.org/0022-3700/8/i=14/a=003}.

\bibitem[{\citenamefont{Yano et~al.}(2014)\citenamefont{Yano, Gao, Goka, and
  Kajita}}]{LightShiftCPT}
\bibinfo{author}{\bibfnamefont{Y.}~\bibnamefont{Yano}},
  \bibinfo{author}{\bibfnamefont{W.}~\bibnamefont{Gao}},
  \bibinfo{author}{\bibfnamefont{S.}~\bibnamefont{Goka}}, \bibnamefont{and}
  \bibinfo{author}{\bibfnamefont{M.}~\bibnamefont{Kajita}},
  \bibinfo{journal}{Phys. Rev. A} \textbf{\bibinfo{volume}{90}},
  \bibinfo{pages}{013826} (\bibinfo{year}{2014}),
  \urlprefix\url{http://link.aps.org/doi/10.1103/PhysRevA.90.013826}.

\bibitem[{\citenamefont{Breschi et~al.}(2010)\citenamefont{Breschi, Kazakov,
  Schori, Di~Domenico, Mileti, Litvinov, and Matisov}}]{LightShiftCPT2}
\bibinfo{author}{\bibfnamefont{E.}~\bibnamefont{Breschi}},
  \bibinfo{author}{\bibfnamefont{G.}~\bibnamefont{Kazakov}},
  \bibinfo{author}{\bibfnamefont{C.}~\bibnamefont{Schori}},
  \bibinfo{author}{\bibfnamefont{G.}~\bibnamefont{Di~Domenico}},
  \bibinfo{author}{\bibfnamefont{G.}~\bibnamefont{Mileti}},
  \bibinfo{author}{\bibfnamefont{A.}~\bibnamefont{Litvinov}}, \bibnamefont{and}
  \bibinfo{author}{\bibfnamefont{B.}~\bibnamefont{Matisov}},
  \bibinfo{journal}{Phys. Rev. A} \textbf{\bibinfo{volume}{82}},
  \bibinfo{pages}{063810} (\bibinfo{year}{2010}),
  \urlprefix\url{http://link.aps.org/doi/10.1103/PhysRevA.82.063810}.

\bibitem[{\citenamefont{Brillet}({1981})}]{BrilletMetrologia}
\bibinfo{author}{\bibfnamefont{A.}~\bibnamefont{Brillet}},
  \bibinfo{journal}{{Metrologia}} \textbf{\bibinfo{volume}{{17}}},
  \bibinfo{pages}{147} (\bibinfo{year}{{1981}}), ISSN
  \bibinfo{issn}{{0026-1394}}.

\bibitem[{\citenamefont{Ohshima et~al.}({1991})\citenamefont{Ohshima, Nakadan,
  Ikegami, and Koga}}]{OshimaCsBeam}
\bibinfo{author}{\bibfnamefont{S.}~\bibnamefont{Ohshima}},
  \bibinfo{author}{\bibfnamefont{Y.}~\bibnamefont{Nakadan}},
  \bibinfo{author}{\bibfnamefont{T.}~\bibnamefont{Ikegami}}, \bibnamefont{and}
  \bibinfo{author}{\bibfnamefont{Y.}~\bibnamefont{Koga}},
  \bibinfo{journal}{{IEEE Transactions On Instrumentation And Measurement}}
  \textbf{\bibinfo{volume}{{40}}}, \bibinfo{pages}{1003}
  (\bibinfo{year}{{1991}}), ISSN \bibinfo{issn}{{0018-9456}}.

\bibitem[{\citenamefont{Budker et~al.}(2005)\citenamefont{Budker, Hollberg,
  Kimball, Kitching, Pustelny, and Yashchuk}}]{BudkerMicrowave}
\bibinfo{author}{\bibfnamefont{D.}~\bibnamefont{Budker}},
  \bibinfo{author}{\bibfnamefont{L.}~\bibnamefont{Hollberg}},
  \bibinfo{author}{\bibfnamefont{D.~F.} \bibnamefont{Kimball}},
  \bibinfo{author}{\bibfnamefont{J.}~\bibnamefont{Kitching}},
  \bibinfo{author}{\bibfnamefont{S.}~\bibnamefont{Pustelny}}, \bibnamefont{and}
  \bibinfo{author}{\bibfnamefont{V.~V.} \bibnamefont{Yashchuk}},
  \bibinfo{journal}{Phys. Rev. A} \textbf{\bibinfo{volume}{71}},
  \bibinfo{pages}{012903} (\bibinfo{year}{2005}),
  \urlprefix\url{http://link.aps.org/doi/10.1103/PhysRevA.71.012903}.

\bibitem[{\citenamefont{Sulai et~al.}(2013)\citenamefont{Sulai, Wyllie, Kauer,
  Smetana, Wakai, and Walker}}]{WalkerACStarkInSERF}
\bibinfo{author}{\bibfnamefont{I.~A.} \bibnamefont{Sulai}},
  \bibinfo{author}{\bibfnamefont{R.}~\bibnamefont{Wyllie}},
  \bibinfo{author}{\bibfnamefont{M.}~\bibnamefont{Kauer}},
  \bibinfo{author}{\bibfnamefont{G.~S.} \bibnamefont{Smetana}},
  \bibinfo{author}{\bibfnamefont{R.~T.} \bibnamefont{Wakai}}, \bibnamefont{and}
  \bibinfo{author}{\bibfnamefont{T.~G.} \bibnamefont{Walker}},
  \bibinfo{journal}{Opt. Lett.} \textbf{\bibinfo{volume}{38}},
  \bibinfo{pages}{974} (\bibinfo{year}{2013}),
  \urlprefix\url{http://ol.osa.org/abstract.cfm?URI=ol-38-6-974}.

\bibitem[{\citenamefont{Podvyaznyi et~al.}({2003})\citenamefont{Podvyaznyi,
  Sakantsev, and Semenov}}]{Podvyaznyi}
\bibinfo{author}{\bibfnamefont{A.}~\bibnamefont{Podvyaznyi}},
  \bibinfo{author}{\bibfnamefont{A.}~\bibnamefont{Sakantsev}},
  \bibnamefont{and} \bibinfo{author}{\bibfnamefont{V.}~\bibnamefont{Semenov}},
  \bibinfo{journal}{{Russian Physics Journal}} \textbf{\bibinfo{volume}{{46}}},
  \bibinfo{pages}{933} (\bibinfo{year}{{2003}}).

\bibitem[{\citenamefont{Camparo et~al.}(1983)\citenamefont{Camparo, Frueholz,
  and Volk}}]{CamparoInhomogeneousLS}
\bibinfo{author}{\bibfnamefont{J.~C.} \bibnamefont{Camparo}},
  \bibinfo{author}{\bibfnamefont{R.~P.} \bibnamefont{Frueholz}},
  \bibnamefont{and} \bibinfo{author}{\bibfnamefont{C.~H.} \bibnamefont{Volk}},
  \bibinfo{journal}{Phys. Rev. A} \textbf{\bibinfo{volume}{27}},
  \bibinfo{pages}{1914} (\bibinfo{year}{1983}),
  \urlprefix\url{http://link.aps.org/doi/10.1103/PhysRevA.27.1914}.

\bibitem[{\citenamefont{Skalla et~al.}({1995})\citenamefont{Skalla, Lang, and
  Wackerle}}]{SkallaLightShift}
\bibinfo{author}{\bibfnamefont{J.}~\bibnamefont{Skalla}},
  \bibinfo{author}{\bibfnamefont{S.}~\bibnamefont{Lang}}, \bibnamefont{and}
  \bibinfo{author}{\bibfnamefont{G.}~\bibnamefont{Wackerle}},
  \bibinfo{journal}{{Journal Of The Optical Society Of America B-optical
  Physics}} \textbf{\bibinfo{volume}{{12}}}, \bibinfo{pages}{772}
  (\bibinfo{year}{{1995}}), ISSN \bibinfo{issn}{{0740-3224}}.

\bibitem[{\citenamefont{Bulos et~al.}({1971})\citenamefont{Bulos, Marshall, and
  Happer}}]{BulosRealTransitions}
\bibinfo{author}{\bibfnamefont{B.}~\bibnamefont{Bulos}},
  \bibinfo{author}{\bibfnamefont{A.}~\bibnamefont{Marshall}}, \bibnamefont{and}
  \bibinfo{author}{\bibfnamefont{W.}~\bibnamefont{Happer}},
  \bibinfo{journal}{{Phys. Rev. A}} \textbf{\bibinfo{volume}{{4}}},
  \bibinfo{pages}{51} (\bibinfo{year}{{1971}}).

\bibitem[{\citenamefont{Cohen-Tannoudji and
  Dupont-Roc}(1972)}]{CohenTannoudjiZeemanLS}
\bibinfo{author}{\bibfnamefont{C.}~\bibnamefont{Cohen-Tannoudji}}
  \bibnamefont{and}
  \bibinfo{author}{\bibfnamefont{J.}~\bibnamefont{Dupont-Roc}},
  \bibinfo{journal}{Phys. Rev. A} \textbf{\bibinfo{volume}{5}},
  \bibinfo{pages}{968} (\bibinfo{year}{1972}),
  \urlprefix\url{http://link.aps.org/doi/10.1103/PhysRevA.5.968}.

\bibitem[{\citenamefont{Higbie et~al.}(2006)\citenamefont{Higbie, Corsini, and
  Budker}}]{Higbie2006}
\bibinfo{author}{\bibfnamefont{J.~M.} \bibnamefont{Higbie}},
  \bibinfo{author}{\bibfnamefont{E.}~\bibnamefont{Corsini}}, \bibnamefont{and}
  \bibinfo{author}{\bibfnamefont{D.}~\bibnamefont{Budker}},
  \bibinfo{journal}{Rev. Sci. Instrum.} \textbf{\bibinfo{volume}{77}},
  \bibinfo{eid}{113106} (\bibinfo{year}{2006}),
  \urlprefix\url{http://scitation.aip.org/content/aip/journal/rsi/77/11/10.1063/1.2370597}.

\bibitem[{\citenamefont{Acosta et~al.}(2006)\citenamefont{Acosta, Ledbetter,
  Rochester, Budker, Jackson~Kimball, Hovde, Gawlik, Pustelny, Zachorowski, and
  Yashchuk}}]{Acosta2006}
\bibinfo{author}{\bibfnamefont{V.}~\bibnamefont{Acosta}},
  \bibinfo{author}{\bibfnamefont{M.~P.} \bibnamefont{Ledbetter}},
  \bibinfo{author}{\bibfnamefont{S.~M.} \bibnamefont{Rochester}},
  \bibinfo{author}{\bibfnamefont{D.}~\bibnamefont{Budker}},
  \bibinfo{author}{\bibfnamefont{D.~F.} \bibnamefont{Jackson~Kimball}},
  \bibinfo{author}{\bibfnamefont{D.~C.} \bibnamefont{Hovde}},
  \bibinfo{author}{\bibfnamefont{W.}~\bibnamefont{Gawlik}},
  \bibinfo{author}{\bibfnamefont{S.}~\bibnamefont{Pustelny}},
  \bibinfo{author}{\bibfnamefont{J.}~\bibnamefont{Zachorowski}},
  \bibnamefont{and} \bibinfo{author}{\bibfnamefont{V.~V.}
  \bibnamefont{Yashchuk}}, \bibinfo{journal}{Phys. Rev. A}
  \textbf{\bibinfo{volume}{73}}, \bibinfo{pages}{053404}
  (\bibinfo{year}{2006}),
  \urlprefix\url{http://link.aps.org/doi/10.1103/PhysRevA.73.053404}.

\bibitem[{\citenamefont{Patton et~al.}(2012)\citenamefont{Patton, Versolato,
  Hovde, Corsini, Higbie, and Budker}}]{Patton2012}
\bibinfo{author}{\bibfnamefont{B.}~\bibnamefont{Patton}},
  \bibinfo{author}{\bibfnamefont{O.~O.} \bibnamefont{Versolato}},
  \bibinfo{author}{\bibfnamefont{D.~C.} \bibnamefont{Hovde}},
  \bibinfo{author}{\bibfnamefont{E.}~\bibnamefont{Corsini}},
  \bibinfo{author}{\bibfnamefont{J.~M.} \bibnamefont{Higbie}},
  \bibnamefont{and} \bibinfo{author}{\bibfnamefont{D.}~\bibnamefont{Budker}},
  \bibinfo{journal}{Applied Physics Letters} \textbf{\bibinfo{volume}{101}},
  \bibinfo{eid}{083502} (\bibinfo{year}{2012}),
  \urlprefix\url{http://scitation.aip.org/content/aip/journal/apl/101/8/10.1063/1.4747206}.

\bibitem[{\citenamefont{Groeger et~al.}(2006)\citenamefont{Groeger, Bison,
  Schenker, Wynands, and Weis}}]{Bison2006}
\bibinfo{author}{\bibfnamefont{S.}~\bibnamefont{Groeger}},
  \bibinfo{author}{\bibfnamefont{G.}~\bibnamefont{Bison}},
  \bibinfo{author}{\bibfnamefont{J.-L.} \bibnamefont{Schenker}},
  \bibinfo{author}{\bibfnamefont{R.}~\bibnamefont{Wynands}}, \bibnamefont{and}
  \bibinfo{author}{\bibfnamefont{A.}~\bibnamefont{Weis}}, \bibinfo{journal}{EPJ
  D} \textbf{\bibinfo{volume}{38}}, \bibinfo{pages}{239}
  (\bibinfo{year}{2006}), ISSN \bibinfo{issn}{1434-6060},
  \urlprefix\url{http://dx.doi.org/10.1140/epjd/e2006-00037-y}.

\bibitem[{\citenamefont{Grujić et~al.}(2015)\citenamefont{Grujić, Koss,
  Bison, and Weis}}]{Bison2015}
\bibinfo{author}{\bibfnamefont{Z.}~\bibnamefont{Grujić}},
  \bibinfo{author}{\bibfnamefont{P.}~\bibnamefont{Koss}},
  \bibinfo{author}{\bibfnamefont{G.}~\bibnamefont{Bison}}, \bibnamefont{and}
  \bibinfo{author}{\bibfnamefont{A.}~\bibnamefont{Weis}}, \bibinfo{journal}{EPJ
  D} \textbf{\bibinfo{volume}{69}}, \bibinfo{eid}{135} (\bibinfo{year}{2015}),
  ISSN \bibinfo{issn}{1434-6060},
  \urlprefix\url{http://dx.doi.org/10.1140/epjd/e2015-50875-3}.

\bibitem[{\citenamefont{Seltzer et~al.}(2007)\citenamefont{Seltzer, Meares, and
  Romalis}}]{Seltzer2007}
\bibinfo{author}{\bibfnamefont{S.~J.} \bibnamefont{Seltzer}},
  \bibinfo{author}{\bibfnamefont{P.~J.} \bibnamefont{Meares}},
  \bibnamefont{and} \bibinfo{author}{\bibfnamefont{M.~V.}
  \bibnamefont{Romalis}}, \bibinfo{journal}{Phys. Rev. A}
  \textbf{\bibinfo{volume}{75}}, \bibinfo{pages}{051407}
  (\bibinfo{year}{2007}),
  \urlprefix\url{http://link.aps.org/doi/10.1103/PhysRevA.75.051407}.

\bibitem[{\citenamefont{Lucivero et~al.}(2014)\citenamefont{Lucivero, Anielski,
  Gawlik, and Mitchell}}]{Lucivero2014}
\bibinfo{author}{\bibfnamefont{V.~G.} \bibnamefont{Lucivero}},
  \bibinfo{author}{\bibfnamefont{P.}~\bibnamefont{Anielski}},
  \bibinfo{author}{\bibfnamefont{W.}~\bibnamefont{Gawlik}}, \bibnamefont{and}
  \bibinfo{author}{\bibfnamefont{M.~W.} \bibnamefont{Mitchell}},
  \bibinfo{journal}{Rev. Sci. Instrum.} \textbf{\bibinfo{volume}{85}},
  \bibinfo{eid}{113108} (\bibinfo{year}{2014}),
  \urlprefix\url{http://scitation.aip.org/content/aip/journal/rsi/85/11/10.1063/1.4901588}.

\bibitem[{\citenamefont{Pustelny et~al.}(2008)\citenamefont{Pustelny,
  Wojciechowski, Gring, Kotyrba, Zachorowski, and Gawlik}}]{Szymek1}
\bibinfo{author}{\bibfnamefont{S.}~\bibnamefont{Pustelny}},
  \bibinfo{author}{\bibfnamefont{A.}~\bibnamefont{Wojciechowski}},
  \bibinfo{author}{\bibfnamefont{M.}~\bibnamefont{Gring}},
  \bibinfo{author}{\bibfnamefont{M.}~\bibnamefont{Kotyrba}},
  \bibinfo{author}{\bibfnamefont{J.}~\bibnamefont{Zachorowski}},
  \bibnamefont{and} \bibinfo{author}{\bibfnamefont{W.}~\bibnamefont{Gawlik}},
  \bibinfo{journal}{Journal of Applied Physics} \textbf{\bibinfo{volume}{103}},
  \bibinfo{eid}{063108} (\bibinfo{year}{2008}),
  \urlprefix\url{http://scitation.aip.org/content/aip/journal/jap/103/6/10.1063/1.2844494}.

\bibitem[{\citenamefont{Balabas et~al.}({2010})\citenamefont{Balabas,
  Karaulanov, Ledbetter, and Budker}}]{supercell}
\bibinfo{author}{\bibfnamefont{M.~V.} \bibnamefont{Balabas}},
  \bibinfo{author}{\bibfnamefont{T.}~\bibnamefont{Karaulanov}},
  \bibinfo{author}{\bibfnamefont{M.~P.} \bibnamefont{Ledbetter}},
  \bibnamefont{and} \bibinfo{author}{\bibfnamefont{D.}~\bibnamefont{Budker}},
  \bibinfo{journal}{{Phys. Rev. Lett.}} \textbf{\bibinfo{volume}{{105}}}
  (\bibinfo{year}{{2010}}), ISSN \bibinfo{issn}{{0031-9007}}.

\bibitem[{\citenamefont{Pustelny et~al.}(2006)\citenamefont{Pustelny,
  Jackson~Kimball, Rochester, Yashchuk, and Budker}}]{SpatialAveraging2006}
\bibinfo{author}{\bibfnamefont{S.}~\bibnamefont{Pustelny}},
  \bibinfo{author}{\bibfnamefont{D.~F.} \bibnamefont{Jackson~Kimball}},
  \bibinfo{author}{\bibfnamefont{S.~M.} \bibnamefont{Rochester}},
  \bibinfo{author}{\bibfnamefont{V.~V.} \bibnamefont{Yashchuk}},
  \bibnamefont{and} \bibinfo{author}{\bibfnamefont{D.}~\bibnamefont{Budker}},
  \bibinfo{journal}{Phys. Rev. A} \textbf{\bibinfo{volume}{74}},
  \bibinfo{pages}{063406} (\bibinfo{year}{2006}),
  \urlprefix\url{http://link.aps.org/doi/10.1103/PhysRevA.74.063406}.

\bibitem[{\citenamefont{Bell and Bloom}(1957)}]{BellBloom}
\bibinfo{author}{\bibfnamefont{W.~E.} \bibnamefont{Bell}} \bibnamefont{and}
  \bibinfo{author}{\bibfnamefont{A.~L.} \bibnamefont{Bloom}},
  \bibinfo{journal}{Phys. Rev.} \textbf{\bibinfo{volume}{107}},
  \bibinfo{pages}{1559} (\bibinfo{year}{1957}),
  \urlprefix\url{http://link.aps.org/doi/10.1103/PhysRev.107.1559}.

\bibitem[{\citenamefont{edited~by Dmitry~{Budker} and
  {Kimball}}(2013)}]{DimaOpticalMagnetometry}
\bibinfo{author}{\bibnamefont{edited~by Dmitry~{Budker}}} \bibnamefont{and}
  \bibinfo{author}{\bibfnamefont{D.~F.~J.} \bibnamefont{{Kimball}}},
  \emph{\bibinfo{title}{Optical Magnetometry}} (\bibinfo{publisher}{Cambridge
  University Press}, \bibinfo{address}{New York}, \bibinfo{year}{2013}), ISBN
  \bibinfo{isbn}{978-1-107-01035-2}.

\bibitem[{\citenamefont{Le~Kien et~al.}(2013)\citenamefont{Le~Kien,
  Schneeweiss, and Rauschenbeutel}}]{Rauschenbeutel}
\bibinfo{author}{\bibfnamefont{F.}~\bibnamefont{Le~Kien}},
  \bibinfo{author}{\bibfnamefont{P.}~\bibnamefont{Schneeweiss}},
  \bibnamefont{and}
  \bibinfo{author}{\bibfnamefont{A.}~\bibnamefont{Rauschenbeutel}},
  \bibinfo{journal}{The European Physical Journal D}
  \textbf{\bibinfo{volume}{67}}, \bibinfo{eid}{92} (\bibinfo{year}{2013}), ISSN
  \bibinfo{issn}{1434-6060},
  \urlprefix\url{http://dx.doi.org/10.1140/epjd/e2013-30729-x}.

\bibitem[{\citenamefont{Budker et~al.}(2008)\citenamefont{Budker, Kimball, and
  DeMIlle}}]{DimaProblems}
\bibinfo{author}{\bibfnamefont{D.}~\bibnamefont{Budker}},
  \bibinfo{author}{\bibfnamefont{D.}~\bibnamefont{Kimball}}, \bibnamefont{and}
  \bibinfo{author}{\bibfnamefont{D.}~\bibnamefont{DeMIlle}},
  \emph{\bibinfo{title}{Atomic physics: An exploration through problems and
  solutions}} (\bibinfo{publisher}{OUP Oxford}, \bibinfo{year}{2008}), ISBN
  \bibinfo{isbn}{9780199532414}.

\bibitem[{\citenamefont{Patton et~al.}(2014)\citenamefont{Patton, Zhivun,
  Hovde, and Budker}}]{vectormag}
\bibinfo{author}{\bibfnamefont{B.}~\bibnamefont{Patton}},
  \bibinfo{author}{\bibfnamefont{E.}~\bibnamefont{Zhivun}},
  \bibinfo{author}{\bibfnamefont{D.~C.} \bibnamefont{Hovde}}, \bibnamefont{and}
  \bibinfo{author}{\bibfnamefont{D.}~\bibnamefont{Budker}},
  \bibinfo{journal}{Phys. Rev. Lett.} \textbf{\bibinfo{volume}{113}},
  \bibinfo{pages}{013001} (\bibinfo{year}{2014}),
  \urlprefix\url{http://link.aps.org/doi/10.1103/PhysRevLett.113.013001}.

\bibitem[{\citenamefont{Corwin et~al.}(1998)\citenamefont{Corwin, Lu, Hand,
  Epstein, and Wieman}}]{DAVLL1}
\bibinfo{author}{\bibfnamefont{K.~L.} \bibnamefont{Corwin}},
  \bibinfo{author}{\bibfnamefont{Z.-T.} \bibnamefont{Lu}},
  \bibinfo{author}{\bibfnamefont{C.~F.} \bibnamefont{Hand}},
  \bibinfo{author}{\bibfnamefont{R.~J.} \bibnamefont{Epstein}},
  \bibnamefont{and} \bibinfo{author}{\bibfnamefont{C.~E.}
  \bibnamefont{Wieman}}, \bibinfo{journal}{Appl. Opt.}
  \textbf{\bibinfo{volume}{37}}, \bibinfo{pages}{3295} (\bibinfo{year}{1998}),
  \urlprefix\url{http://ao.osa.org/abstract.cfm?URI=ao-37-15-3295}.

\bibitem[{\citenamefont{Lee et~al.}(2011)\citenamefont{Lee, Iwata, Corsini,
  Higbie, Knappe, Ledbetter, and Budker}}]{DAVLL2}
\bibinfo{author}{\bibfnamefont{C.}~\bibnamefont{Lee}},
  \bibinfo{author}{\bibfnamefont{G.~Z.} \bibnamefont{Iwata}},
  \bibinfo{author}{\bibfnamefont{E.}~\bibnamefont{Corsini}},
  \bibinfo{author}{\bibfnamefont{J.~M.} \bibnamefont{Higbie}},
  \bibinfo{author}{\bibfnamefont{S.}~\bibnamefont{Knappe}},
  \bibinfo{author}{\bibfnamefont{M.~P.} \bibnamefont{Ledbetter}},
  \bibnamefont{and} \bibinfo{author}{\bibfnamefont{D.}~\bibnamefont{Budker}},
  \bibinfo{journal}{Rev. Sci. Instrum.} \textbf{\bibinfo{volume}{82}},
  \bibinfo{eid}{043107} (\bibinfo{year}{2011}),
  \urlprefix\url{http://scitation.aip.org/content/aip/journal/rsi/82/4/10.1063/1.3568824}.

\bibitem[{\citenamefont{Chalupczak et~al.}(2012)\citenamefont{Chalupczak,
  Godun, Anielski, Wojciechowski, Pustelny, and Gawlik}}]{Szymek}
\bibinfo{author}{\bibfnamefont{W.}~\bibnamefont{Chalupczak}},
  \bibinfo{author}{\bibfnamefont{R.~M.} \bibnamefont{Godun}},
  \bibinfo{author}{\bibfnamefont{P.}~\bibnamefont{Anielski}},
  \bibinfo{author}{\bibfnamefont{A.}~\bibnamefont{Wojciechowski}},
  \bibinfo{author}{\bibfnamefont{S.}~\bibnamefont{Pustelny}}, \bibnamefont{and}
  \bibinfo{author}{\bibfnamefont{W.}~\bibnamefont{Gawlik}},
  \bibinfo{journal}{Phys. Rev. A} \textbf{\bibinfo{volume}{85}},
  \bibinfo{pages}{043402} (\bibinfo{year}{2012}),
  \urlprefix\url{http://link.aps.org/doi/10.1103/PhysRevA.85.043402}.

\bibitem[{\citenamefont{Budker et~al.}({2002})\citenamefont{Budker, Gawlik,
  Kimball, Rochester, Yashchuk, and Weis}}]{DimaAlignment}
\bibinfo{author}{\bibfnamefont{D.}~\bibnamefont{Budker}},
  \bibinfo{author}{\bibfnamefont{W.}~\bibnamefont{Gawlik}},
  \bibinfo{author}{\bibfnamefont{D.}~\bibnamefont{Kimball}},
  \bibinfo{author}{\bibfnamefont{S.}~\bibnamefont{Rochester}},
  \bibinfo{author}{\bibfnamefont{V.}~\bibnamefont{Yashchuk}}, \bibnamefont{and}
  \bibinfo{author}{\bibfnamefont{A.}~\bibnamefont{Weis}},
  \bibinfo{journal}{{Rev. Mod. Phys.}} \textbf{\bibinfo{volume}{{74}}},
  \bibinfo{pages}{1153} (\bibinfo{year}{{2002}}), ISSN
  \bibinfo{issn}{{0034-6861}}.

\end{thebibliography}

\end{document}